\documentclass[12pt]{article}
\def \be {\begin{equation}}
\def \ee {\end{equation}}
\def \bea {\begin{eqnarray}}
\def \eea {\end{eqnarray}}
\def \nn {\nonumber}

\def \del {\partial}
\def \dels {\partial\kern-.5em / \kern.5em}
\def \As {{A\kern-.5em / \kern.5em}}
\def \Ds {D\kern-.7em / \kern.5em}

\def \d {\delta}

\def \lam {\lambda}

\def \om {\omega}
\def \Om {\Omega}
\def \th {\theta}

\def \cO {{\cal O}}
\def \bast {\bar{\ast}}

\begin{document}

\begin{titlepage}

\begin{center}

\hfill hep-th/0111160\\

\vskip .5in

\textbf{\Large
Perturbative Approach to \\
Higher Derivative and Nonlocal Theories
}

\vskip .5in
{\large Tai-Chung Cheng, Pei-Ming Ho, Mao-Chuang Yeh}
\vskip 15pt

{\small
Department of Physics, \\
National Taiwan University, \\
Taipei 106, Taiwan, \\
R.O.C.}\\

\vskip .2in

\sffamily{pmho@phys.ntu.edu.tw}

\vspace{60pt}

\end{center}

\begin{abstract}

We review 
a perturbative approach to
deal with Lagrangians with higher or
infinite order time derivatives.
It enables us to construct a consistent
Poisson structure and Hamiltonian
with only first time derivatives
order by order in coupling.
We show that, 
to the lowest order,
the Hamiltonian is bounded from below
whenever the potential is.
We consider spacetime noncommutative field theory
as an example.

\end{abstract}


\end{titlepage}

\setcounter{footnote}{0}

\section{Introduction}

In the canonical formulation,
we usually consider Lagrangians
with only first time derivatives.
However, higher derivative theories,
including nonlocal theories,
also have many physical applications.
For example,
when one integrates out
high energy degrees of freedom
in a local field theory,
the low energy effective action
is generically nonlocal \cite{BVDM}.
Higher derivative theories were
considered even earlier in order to
find a finite quantum field theory \cite{nonlocal},
before the advent of renormalization.
Moreover, theories with infinite derivatives
are inevitable from the viewpoint of string theory
\cite{EW,HH}.
There are many other examples,
such as higher derivative gravity \cite{higherderiv},
meson-nucleon interactions \cite{meson},
and spacetime noncommutative field theory \cite{SW,SST1},
and so on.

Unfortunately, an exact treatment of
generic nonlocal theories is so far unknown.
This was viewed as a serious problem for string theory \cite{EW}.
However, in most cases,
higher derivative terms appear as
higher order corrections
in the effective Lagrangian,
so a perturbative approximation scheme would
already be very useful.

In this paper we consider theories
for which the free part of the Lagrangian
involves only first time derivatives,
but there can be higher derivatives
to any (finite or infinite) order
in the interaction terms.
According to the canonical formulation,
the phase space is $2k$ dimensional
for Lagrangians with $k$-th time derivatives
\cite{Ostro}.
For a low energy effective theory, however,
we usually assume that
there are as many degrees of freedom
as the free part of the Lagrangian tells us.
This is also an assumption for the S-matrix to be unitary
in a quantum field theory \cite{Weinb}.
Presumably the extra dimensions of the phase space
are associated with certain high energy degrees of freedom
which are integrated out from the description.

What we need is a formalism 
to deal with
Lagrangians with finite or infinite derivatives 
without
introducing new (high energy) degrees of freedom.
It should also allow 
us to consistently
construct Poisson brackets and Hamiltonian
perturbatively order by order,
and in the end 
give 
us an effective Lagrangian
with only first time derivatives.

In \cite{BG1}, it was 
proposed to remove higher derivative terms in the Lagrangian
by field redefinition.
But it is unclear how to achieve this
to arbitrary orders for a generic theory,
or whether this is always possible.
A general procedure which provides 
a Hamiltonian formulation order by order 
in the number of time derivatives in the interaction terms 
was constructed in \cite{JLM}. 
The perturbative approach \cite{EW} that 
we will discuss in this paper is a modification of that. 
It is based on an expansion in the coupling constant, 
with no restriction to the number of derivatives in each order. 
This technique has been used extensively in \cite{Simon} 
for higher derivative gravitational theory. 
Although this approach was studied in detail in \cite{EW}, 
which we recommend the interested readers to refer to 
for related issues, such as the quantization or causality problems, 
we present this approach here in a slightly different fashion, 
with some derivations and proof explicitly given. 
In fact, we have discovered the same techniques independently, 
and were only acquainted with previous works 
on this subject by a referee after this work was finished. 

What is the new progress we made on this subect? 
(1) We show that, 
to the lowest order, we can describe the result
most simply by a change of variable.
Remarkably,
it turns out that, in terms of this new variable, 
the Hamiltonian is bounded from below
whenever the potential is.
(2) We give the formal proof that the perturbative approach 
can be carried out to all orders consistently. 
This proof also shows that 
if an exact solution to the equation of motion is known, 
the same approach can immediately give 
the final result. 
(3) In order to understand the difference between 
the original system and its perturbative formulation, 
we demonstrate by an example that 
when the calculation is carried out to all orders, 
the description of low energy modes is exact, 
including the effect of their interactions with high energy modes, 
but high energy degrees of freedom are still absent. 
(4) We apply this approach to field theories 
living on noncommutative spacetime, 
and comment on the spacetime uncertainty relation. 

This paper is organized as follows. 
In Sec. \ref{Canonical},
we briefly review the canonical formulation
of theories with finite time derivatives,
and explain the problems in dealing with
infinite derivatives.
We comment there and show in the appendix
that one-loop unitarity
is not affected by introducing
higher spatial derivatives,
but is at stake when
higher time derivatives are present.
In Sec. \ref{Perturb}
the perturbative approach for
higher derivative theories is constructed.
One can have a consistent
Hamiltonian formulation of the theory
to an arbitrary order in coupling.
We also give a formal proof that
this procedure can be consistently
carried out to all orders.
Finally, in Sec. \ref{Examples}
we give three examples.
In the first example \cite{EW}
we demonstrate explicitly
how the perturbative approach works
to the second order.
The second example
is used to show that this approach
suppresses high energy degrees of freedom,
but provides an exact description for the low energy modes.
We also apply this approach to the example of
spacetime noncommutative field theory.

\section{Canonical Formulation} \label{Canonical}

\subsection{Lagrangians with Finite Time Derivatives}

Here we review the canonical formulation
for classical theories with finite time derivatives.
Consider a Lagrangian with $k$-th time derivatives
\begin{equation}
L_0=L_0(q,\dot{q},\ddot{q},\cdots,q^{(k)}),
\end{equation}
where $q^{(i)}$ is the $i$-th time derivative of $q$.
Let us apply the formalism of \cite{FJ}.
The variation of the action
\[
S=\int_{t_i}^{t_f} dt \; L
\]
with respect to $q$
is found to be the time integral of
$\d q$ times the Euler-Lagrange equation
\be \label{EOM0}
\sum_{n=0}^{k} \left(-\frac{d}{dt}\right)^n
\frac{\del L_0}{\del q^{(n)}} = 0,
\ee
plus a boundary term
\be
\left[ \sum_i P_i \d q^{(i)} \right]_{t_i}^{t_f},
\ee
where $P_i$ is the conjugate momentum of $q^{(i)}$
\be \label{Pi}
P_i=
\sum_{j=0}^{k-i-1}\left(-\frac{d}{dt}\right)^j
\frac{\del L_0}{\del q^{(i-j+1)}}.
\ee
The symplectic structure can be directly
read off from this as
\be \label{Om0}
\Om=\sum_{i=0}^{k-1} dP_i d q^{(i)}.
\ee

The Hamiltonian is
\be
H=\sum_{i=0}^{k-1} P_i q^{(i)} -
L(q,\dot{q},\cdots,q^{(k-1)},q^{(k)}),
\ee
where $q^{(k)}$ is viewed
as a function of $q^{(i)}$'s for $i=0,\cdots,k-1$ and $P_{k-1}$.
\footnote{
We assume that the Lagrangian is nondegenerate.
This can always be achieved by
adding total derivatives.
}
Since $H$ is linear in $P_1, \cdots, P_{k-2}$,
the energy spectrum is unbounded from below.

Another approach to deal with higher derivative
Lagrangians is to introduce new variables
to replace higher derivatives of $q$,
and use Lagrange multipliers to ensure that
we have not introduced any new degrees of freedom.
One natural choice is to define new variables $q_i$
by $q_i=q^{(i)}$ (Exercise 1.26 in \cite{HT}),
and another is to set $q_i=q^{(2i)}$.
In any case, after introducing new variables,
the Lagrangian only has first time derivatives.
Canonical momenta of all variables,
including the Lagrange multipliers,
can be defined as usual.
Some of these will impose primary constraints.
Dirac's procedure for constrained quantization
can then be used to obtain the same result.
See \cite{HT} for a detailed description
of the general procedure.

\subsection{Lagrangians with Infinite Time Derivatives}

For a Lagrangian with infinite time derivatives
$L_0(q,\dot{q},\cdots)$,
one may hope that a trick similar to the one
for Lagrangians with finite derivatives will work.
A natural guess is to define the new Lagrangian as
\cite{LV}-\cite{Bering}
\be \label{LQ}
L=L_0(Q)+\int d\lam\; \mu(\lam)(\dot{Q}(\lam)-Q'(\lam)),
\ee
where $L_0(Q)$ is obtained from $L_0[q]$
by the replacement
\be
q(t)\rightarrow Q(t,0),
\quad q^{(n)}(t)\rightarrow
\frac{d^n}{d\lam^n}Q(t,\lam)|_{\lam=0}.
\ee
By $Q'$ we mean $\frac{d}{d\lam}Q$ and
$\mu$ is a Lagrange multiplier.
This implies that $Q(t,\lam)=q(t+\lam)$.

However, the Euler-Lagrange equations
for this new Lagrangian $L$ is in fact
different from the Eular-Lagrange equation
for the original Lagrangian $L_0$.
Further constraints, or boundary conditions,
have to be imposed \cite{LV}-\cite{Bering}.

Instead of trying to resolve this problem,
we give some general reasons why
a naive extension of the idea
for finite time derivatives has difficulties.

For a nondengerate Lagrangian with $k$-th time derivatives,
the phase space is $2k$ dimensional
for a single variable $q$.
Let us choose $q^{(i)}$ $(i=0,1,\cdots,2k-1)$
to be the coordinates on the phase space.
The symplectic 2-form (\ref{Om0}) implies that
the Poisson bracket
\be \label{zero}
(q^{(i)}, q^{(j)})=0
\ee
for any $i,j\leq k-1$.
The bracket is nontrivial
when $i\leq k-1$ and $j\geq k$ and vise versa.

When we take $k\rightarrow\infty$, naively,
the phase space has coordinates
$\{q^{(i)}\}_{i=0,1,\cdots,\infty}$,
and (\ref{zero}) holds for any $i$ and $j$.
The nontrivial part of Poisson brackets is now completely lost!

On the other hand, if we use $q^{(i)}$ and $P_i$
($i=0,1,\cdots,k-1$) as the phase space variables,
in the limit $k\rightarrow\infty$,
the phase space has coordinates
$\{q^{(i)},P_i\}_{i=0,1,\cdots,\infty}$,
which seems to have twice the dimension
of the previous guess.

Another problem with infinite derivatives
is the Cauchy problem \cite{BNW}.
For a Lagrangian with $k$-th derivative
its equation of motion is
a differential equation of order $2k$.
To determine the trajectory uniquely,
the initial values of $q^{(i)}$
for $i=0,1,\cdots,2k-1$, have to be given.
Naively, when $k\rightarrow\infty$,
all $q^{(i)}$'s have to be given as initial conditions,
but they will completely determine the trajectory of $q$
via Taylor expansion even without
using the equation of motion.
This seems to suggest that for infinite $k$
the Euler Lagrange equation should be imposed
as a constraint on the space of initial conditions,
which is equivalent to the phase space,
and the evolution for a point in the phase space
is trivial.
This viewpoint is consistent with the vanishing
of Poisson brackets mentioned above,
and is similar to the proposal of
earlier works \cite{LV}-\cite{Bering}.

Although it is desirable to
have a general formalism to deal with
theories with infinite derivatives.
It is known that different cases
can have very different properties.
For some interesting examples see
\cite{EW},\cite{Woodard}.
Probably different theories
with infinite derivatives may 
have to be treated very differently.

%

Normally, one needs to understand
the classical Hamiltonian formulation of a system
before quantization can be done.
In the operator formalism
we replace the Poisson brackets
by commutators.
In the path integral formalism,
we should start with
\be
\int Dq\; Dp \; e^{\int dt (p\dot{q}-H)}
\ee
for a system with unconstrained
conjugate variables $p$, $q$.
The formula
\be \label{PIL}
\int Dq e^{\int dt \; L}
\ee
may or may not be correct.

Given a Lagrangian with infinite derivatives,
if one simply use (\ref{PIL})
to define a quantum system\footnote{
The Hamiltonian formulation of \cite{GKL}
also leads to (\ref{PIL}).
},
the violation of unitarity is observed
in several examples \cite{BR}.
It is tempting to interpret
$Q(t,\lam)$ in (\ref{LQ}) as a chiral field
on an open string,
and an intriguing possibility is that
for certain inifinite derivative theories
one can find a procedure to add new degrees of freedom
for consistency which eventually leads to string theory.

In the next section we will review 
a perturbative approach
for low energy effective theories
with infinite time derivatives.
This approach allows us to change the Lagrangian
order by order to a new Lagrangian with
only first time derivatives.
It is then possible to quantize
the theory without breaking unitarity.
In the appendix we also show that,
with the free part of the Lagrangian unchanged,
arbitrary spatial derivatives in the interaction terms
will not spoil unitarity at one-loop level,
which is sufficient for low energy effective theories.
In particular, an immediate consequence
is that noncommutative field theories
with only spatial noncommutativity
is unitary at one loop level.

\section{Perturbative Approach} \label{Perturb}

In this section we 
review an approach 
to deal with higher derivative terms perturbatively.
This approach works equally well for Lagrangians
with finite or infinite order time derivatives.

In many applications, higher derivative terms
appear in the Lagrangian
as higher order corrections to
a low energy effective theory.
However, according to the canonical Hamiltonian formulation,
when higher order terms are introduced,
the dimension of the phase space increases,
no matter how small the correction term is.
Intuitively, this discrete jump in dimension
does not seem to be compatible with
the physical interpretation of
the higher derivative interaction terms
as a perturbative correction to the theory.

To resolve this apparent conceptual conflict,
we note that the new dimensions of the phase space
correspond to those degrees of freedom
which are not accessible at low energies.
If all we want is to have some knowledge about
how these high energy states affect
the behavior of the low energy modes,
we can choose to restrict ourselves
to the low energy degrees of freedom on the phase space.
Therefore, the main idea behind 
this 
approach is to
project the symplectic structure on the whole phase space
onto the subspace of low energy states,
and then induce the Hamiltonian formulation
for the reduced phase space.
This approach can also be generalized
to cases in which it is preferred to
work on a bigger phase space with
finitely many higher time derivatives kept in the end.

It is sufficient to illustrate 
this 
approach by considering a $0+1$ dimensional theory
with a general Lagrangian
\be
L=\frac{1}{2}\dot{q}^2-\frac{m}{2}q^2-g
V(q,\dot{q},\ddot{q},\cdots).
\ee
To be definite, let us consider the cases
with infinite order time derivatives.
Under variation, the action is
\be \label{dS}
\d S=-\int dt(\mbox{EOM})\d q +
\left[\sum_{k=0}^{\infty}P_k\d
q^{(k)}\right]_{t_i}^{t_f},
\ee
where the equation of motion is
\be
\label{EOM}
\mbox{EOM}\equiv\ddot{q}+m^2 q+g\sum_{n=0}^{\infty}
\left(-\frac{d}{dt}\right)^n\frac{\del V}{\del q^{(n)}} = 0,
\ee
and $P_n$ is the canonical momenta for $q^{(n)}$
\be
\label{Pn}
P_k=\dot{q}\d_{k0} - g\sum_{n=k+1}^{\infty}
\left(-\frac{d}{dt}\right)^{n-k-1} \frac{\del V}{\del q^{(n)}}.
\ee
The symplectic two-form is
\be
\label{Omega}
\Om=\sum_{k=0}^{\infty}dP_k dq^{(k)}.
\ee

\subsection{First Order Approximation} \label{FirstOrder}

In an exact treatment,
all $q^{(n)}$'s are independent.
In the low energy approximation,
we only keep $q$ and $\dot{q}$.
Using the equation of motion,
we can replace higher derivatives of $q$
by lower derivatives.
To the lowest order approximation,
\be
\label{qn}
q^{(n)}\approx\left\{\begin{array}{ll}
                       (-m^2)^{n/2}q, & (n=\mbox{even}), \\
                       (-m^2)^{(n-1)/2}\dot{q} & (n=\mbox{odd}).
                     \end{array}\right.
\ee
The boundary term of $\d S$ (\ref{dS}) reduces to
\be
\left[\Pi_0\d q+\Pi_1\d\dot{q}\right]_{t_i}^{t_f}=
\left[(\dot{q}-g\xi_0)\d q-g\xi_1\d\dot{q}\right]_{t_i}^{t_f},
\ee
where
\bea
\xi_0&=&\sum_{k=0}^{\infty}\sum_{n=2k+1}^{\infty}
(-m^2)^k\left(-\frac{d}{dt}\right)^{n-2k-1}
\frac{\del V}{\del q^{(n)}}, \\
\xi_1&=&\sum_{k=0}^{\infty}\sum_{n=2k+2}^{\infty}
(-m^2)^k\left(-\frac{d}{dt}\right)^{n-2k-2}
\frac{\del V}{\del q^{(n)}}.
\eea
Note that $\xi_0$ and $\xi_1$ should be viewed as
functions of $q$ and $\dot{q}$ via the replacement (\ref{qn}).

We find the symplectic two-form (\ref{Omega}) as
\be
\label{Om1}
\Om=d\Pi_0 dq+d\Pi_1 d\dot{q}=
\left(-1+g\frac{\del\xi_0}{\del\dot{q}}-
g\frac{\del\xi_1}{\del q}\right)dq d\dot{q}.
\ee
It follows that the Poisson bracket should be
\be \label{qqdot}
(q, \dot{q})\simeq
1+g\frac{\del\xi_0}{\del\dot{q}}-g\frac{\del\xi_1}{\del q}.
\ee


To the lowest order in $g$,
the Poisson bracket can be written as
\be
(x, p)=1,
\ee
where
\be
x=q+g\xi_1, \quad p=\dot{q}-g\xi_0.
\ee

The Hamiltonian is
\be \label{Ham}
H=\Pi_0\dot{q}+\Pi_1\ddot{q}-L,
\ee
where we replace all higher derivatives of $q$
by (\ref{qn}).
We find
\be
H=\frac{1}{2}p^2+\frac{m^2}{2}x^2+g\tilde{V}(x,p),
\ee
where $\tilde{V}(x,p)=V(x,p,-m^2 x,\cdots)$.
Note that, if the potential $V$ is bounded from below,
the first order Hamiltonian is also bounded from below.

One can check that the Hamilton equations
\be \label{HE}
\dot{q}=(q, H), \quad \ddot{q}=(\dot{q}, H)
\ee
reproduce the equation of motion (\ref{EOM})
to the first order in $g$.
The corresponding Lagrangian is
\be \label{LL}
\tilde{L}=p\dot{x}-H=
\frac{1}{2}\dot{x}^2-\frac{m^2}{2}x^2-g\tilde{V}(x,\dot{x}).
\ee
Its Euler-Lagrange equation agrees with
the original system to the first order in $g$.
Since the final expression only contains
first derivatives,
its quantization is straightforward.

\subsection{Higher Order Approximation}

For higher order corrections,
we first iterate the equation of motion (\ref{EOM})
up to a certain order $\cO(g^n)$.
For example, at the first order,
\be \label{q2}
q^{(2)}\rightarrow -m^2 q-g\sum_{n=0}^{\infty}
\left(-\frac{d}{dt}\right)^n\frac{\del V}{\del q^{(n)}},
\ee
where we should also use the replacement (\ref{qn})
to change the last term into a function of $q$ and $\dot{q}$ only.
Higher derivatives of $q$ can also be
replaced by functions of $q$ and $\dot{q}$
up to the same order in $g$ by
differentiating with respect to time
and repeatedly using (\ref{q2}) as
\be
\label{qn2}
q^{(n)}\approx\left\{
\begin{array}{ll}
(-m^2)^{n/2}q-g\sum_{l=1}^{n/2}\sum_{k=0}^{\infty}
(-m^2)^{n/2-l}\left(\frac{d}{dt}\right)^{k+l-1} \frac{\del V}{\del
q^{(k)}} & (n=\mbox{even}), \\
(-m^2)^{(n-1)/2}\dot{q}-g\sum_{l=1}^{(n-1)/2}\sum_{k=0}^{\infty}
(-m^2)^{(n-1)/2-l}\left(\frac{d}{dt}\right)^{k+l-1} \frac{\del
V}{\del q^{(k)}} & (n=\mbox{odd}).
\end{array}\right.
\ee

In general, we can have all $q^{(i)}$'s
expressed as a function of $q$ and $\dot{q}$ only,
up to a certain order ${\cal O}(g^n)$.
This helps us to derived the symplectic form
from (\ref{Omega})
\be \label{sympl}
\Om=\sum_i dP_i dq^{(i)}=d\Pi_0 dq+d\Pi_1 d\dot{q}
=\left(-\frac{\del\Pi_0}{\del\dot{q}}+
\frac{\del\Pi_1}{\del q}\right)dq d\dot{q}.
\ee

The final Hamiltonian is defined by (\ref{Ham})
with all higher derivatives of $q$ replaced by
functions of $q$, $\dot{q}$.
Its Hamilton equations (\ref{HE})
gives the equation of motion up to $\cO(g^{n+1})$.

An effective Lagrangian can be defined as
\be
\tilde{L}=p\dot{x}-H
\ee
for a conjugate pair $(x, p)=1$.
For such Lagrangians
we do not expect violation of unitarity
since it contains only first time derivatives.
Although we expect such theories to be
nonrenormalizable in the usual (power-counting) sense,
it remains to be seen whether it is renormalizable
in the mordern sense \cite{GW}.

\subsection{To All Orders: A Formal Proof}

Now we give a formal proof for the self-consistency
of the 
perturbative formulation.
{}From the equation of motion (\ref{EOM}),
in principle one can find
\be \label{qf}
\ddot{q}=f(q,\dot{q})
\ee
for a certain function $f$ to all orders in $g$
by iteration (or inspiration).
Higher derivatives of $q$ can be derived
from this
\be \label{qfn}
q^{(n)}=f_n(q,\dot{q}),
\ee
where the functions $f_n$ can
be obtained recursively
\be
f_{n+1}=\left[\frac{d}{dt}f_n\right]=
\frac{\del f_n}{\del q}\dot{q}+
\frac{\del f_n}{\del\dot{q}}f,
\ee
where we used the notation
\be
[A]\equiv A|_{q^{(n)}=f_n}.
\ee

Here are a few identities that
will come in handy in the following.
From (\ref{Pn}), we find
\be \label{id1}
\dot{P}_k=\frac{\del L_0}{\del q^{(k)}}-P_{k-1}.
\ee
For an arbitrary function $A$ on the phase space,
we also find the following identities
\bea
\frac{d}{dt}[A]&=&\left[\dot{A}\right]+
\frac{\del [A]}{\del\dot{q}}(\ddot{q}-f), \label{id2} \\
\frac{\del [A]}{\del q}&=&
\sum_{k=0}^{\infty}\left[\frac{\del A}{\del q^{(k)}}\right]
\frac{\del f_k}{\del q}, \label{id3}
\eea
where one can also replace
$\frac{\del}{\del q}$ by $\frac{\del}{\del\dot{q}}$
in the last formula.

The variables $\Pi_0$, $\Pi_1$ can be read off from
\be
\Pi_0\d q+\Pi_1\d\dot{q}=\sum_{k=0}^{\infty}P_k\d q^{(k)},
\ee
and we find
\bea
\Pi_0&=
&\sum_{k=0}^{\infty}
\left[ P_k\frac{\del q^{(k)}}{\del q}\right], \label{Pi0}\\
\Pi_1&=&\sum_{k=0}^{\infty}
\left[ P_k\frac{\del q^{(k)}}{\del\dot{q}}\right]. \label{Pi1}
\eea
The Hamiltonian is
\be
H=[\Pi_0\dot{q}+\Pi_1\ddot{q}-L].
\ee
The Hamilton equations (\ref{HE}) based on
the symplectic structure (\ref{sympl}) are
\bea
&\dot{\Pi}_0+\Pi_1\frac{\del f}{\del q}
-\frac{\del \Pi_1}{\del q}(\ddot q-f)
-\frac{\del[L_0]}{\del q}=0, \label{eq1} \\
&\dot{\Pi}_1+\Pi_0+\Pi_1\frac{\del f}{\del \dot{q}}
-\frac{\del\Pi_1}{\del\dot{q}}(\ddot q-f)
-\frac{\del[L_0]}{\del\dot{q}}=0. \label{eq2}
\eea
With the help of (\ref{id1})-(\ref{id3}),
one can show from (\ref{Pi0}), (\ref{Pi1})
that (\ref{eq2}) is automatically satisfied,
and that (\ref{eq1}) is equivalent to
the equation of motion (\ref{qf}).

Certainly, (\ref{qf}) is not the only solution
of the equation of motion (\ref{EOM}).
It may be extended to a more general solution
with many free parameters
\be
\ddot{q}=f(q,\dot{q},c_1,c_2,\cdots).
\ee
These extra parameters are associated with
high energy degrees of freedom.
Ignoring them, or setting them to zeros
corresponds to taking the low energy limit
in which these high energy modes are
``minimally excited''.
That is, they are only excited to be
consistent with the equation of motion,
but we do not consider their independent
degrees of freedom.
We will consider a system of coupled springs
in the next section as an example
to show this property more concretely.

\section{Examples} \label{Examples}

\subsection{An Example of Second Order Approximation}

In this section we show the perturbative procedure
explicitly for an example to the second order.
Consider the Lagrangian \cite{EW}
\be
L=\frac{1}{2}\dot{q}^2+\frac{a}{2}q^2+gq\ddot{q}^2.
\ee
Its Euler-Lagrange equation is
\be \label{EOMq}
\ddot{q}=aq+g\ddot{q}^2+2g(q\ddot{q})^{\cdot\cdot}.
\ee

Following the procedure outlined in the previous section,
we find
\be
\Pi_0=\dot{q}-2g(q\ddot{q})^{\cdot},
\quad \Pi_1=2g q\ddot{q}
\ee
from (\ref{Pi0}), (\ref{Pi1}).
Iterating (\ref{EOMq}) to the first order in $g$, we find
\be
\ddot{q}=
aq+g(5a^2 q^2+4a\dot{q}^2)+\cO(g^2).
\label{EOMq2}
\ee
Taking derivatives, we get
\bea
q^{(3)}&=&a\dot{q}+g(18a^2q\dot{q})+\cO(g^2), \nn\\
q^{(4)}&=&a^2 q+g(23a^3 q^2+22a^2\dot{q}^2)+\cO(g^2),
\eea
and so on.

Up to third order terms in $g$, we find
\bea
\Pi_0&=&\dot{q}
-4g(aq\dot{q})-2g^2(23a^2 q^2\dot{q}+4a\dot{q}^3), \\
\Pi_1&=&
2g(aq^2)+2g^2(5a^2 q^3+4aq\dot{q}^2).
\eea
The Poisson bracket is
\be
(q,\dot{q})
=1+8gaq+4g^2(35a^2 q^2+8a\dot{q}^2)+\cO(g^3).
\label{qqdot1}
\ee
The Hamiltonian is
\be
H=\frac{1}{2}\dot{q}^2-\frac{a}{2}q^2+g(a^2 q^3-4aq\dot{q}^2)
+2g^2(5a^3 q^4-19a^2 q^2\dot{q}^2-4a\dot{q}^4)+\cO(g^3).
\ee
One can check that the Hamilton equations (\ref{HE})
give the equation of motion (\ref{EOMq})
correct to the third order in $g$.

The effective Lagrangian is
\be
\tilde{L}=p\dot{q}-H,
\ee
where $p$ is the phase space variable conjugate to $q$,
i.e., $(q,p)=1$.
{}From (\ref{qqdot1}), we can solve for $p$
\be
p=\dot{q}-8gaq\dot{q}-
4g^2(19a^2 q^2\dot{q}+\frac{8}{3}a\dot{q}^3)
+\cO(g^3).
\ee
Thus we find the Lagrangian for $q, \dot{q}$
\be
\tilde{L}=\frac{1}{2}\dot{q}^2
+\frac{a}{2}q^2-g(a^2 q^3+4aq\dot{q}^2)
+2g^2(-5a^3 q^4-19a^2
q^2\dot{q}^2-\frac{4}{3}a\dot{q}^4)+\cO(g^3).
\ee
The Euler-Lagrange equation derived from this Lagrangian is
\be
\ddot{q}=aq+g(5a^2 q^2+4a\dot{q}^2)
+4g^2(19a^3 q^3+35a^2
q\dot{q}^2)+\cO(g^3).
\ee
This is exactly what one gets from (\ref{EOMq})
by iteration, in the same way we got (\ref{EOMq2}),
but to a higher order correction in $g$.


\subsection{Coupled Springs}

Consider two springs $A$ and $B$
of spring constants $k$ and $K$, respectively,
as in Fig. 1.
One end of $A$ is fixed to a wall.
The other end of $A$ is connected to $B$,
with a mass $M$ attached to the joint.
Another mass $m$ is attached to the end of $B$.

{\center
\begin{picture}(250,120)(-200,0) \label{Fig1}
\put(-30,100){\line(1,0){60}}
\put(0,100){\line(0,-1){5}}
\put(0,95){\line(1,0){5}}
\put(5,95){\line(-2,-1){10}}
\put(-5,90){\line(1,0){10}}
\put(5,90){\line(-2,-1){10}}
\put(-5,85){\line(1,0){10}}
\put(5,85){\line(-2,-1){10}}
\put(-5,80){\line(1,0){10}}
\put(5,80){\line(-2,-1){10}}
\put(-5,75){\line(1,0){5}}
\put(0,75){\line(0,-1){10}}
\put(0,70){\circle*{6}}
\put(0,65){\line(1,0){5}}
\put(5,65){\line(-2,-1){10}}
\put(-5,60){\line(1,0){10}}
\put(5,60){\line(-2,-1){10}}
\put(-5,55){\line(1,0){10}}
\put(5,55){\line(-2,-1){10}}
\put(-5,50){\line(1,0){10}}
\put(5,50){\line(-2,-1){10}}
\put(-5,45){\line(1,0){5}}
\put(0,45){\line(0,-1){5}}
\put(0,40){\circle*{6}}
\put(-20,10){Figure 1}
\end{picture}
}

For $K\gg k$ and $M\sim m$,
the degrees of freedom of $M$
is much harder to excite than $m$.
Let the coordinates of $m$ and $M$ be $x$ and $y$.
The equations of motion are
\bea
&m\ddot{x}+k(x-y)=0, \\
&M\ddot{y}+Ky+k(y-x)=0.
\eea
They imply that
\be
ax^{(4)}+b\ddot{x}+cx=0,
\ee
where $a=Mm$, $b=(M+m)k+mK$, $c=Kk$.
The effective Lagrangian for this equation of motion is
\be \label{eL}
L\propto\frac{1}{2}(-a{\ddot{x}}^2+b\dot{x}^2-cx^2).
\ee

Assume that $K\gg k$, then
the first term is negligible for small $\ddot{x}$.
Imagine that someone studies
the behavior of the spring system at low energies
by conducting experiments in which only
the variable $x$ is manipulated,
and finds an approximate Lagrangian as (\ref{eL})
without the first term.
The prediction of the natural frequence is
\be \label{badone}
\om_0^2=\frac{Kk}{(M+m)k+mK}\simeq
\frac{k}{m}\left(1-\frac{k}{K}\frac{M+m}{m}\right)
\ee
to the lowest order in $\cO(k/K)$.

Now suppose that via more accurate experiments,
the first term in the Lagrangian (\ref{eL}) is put in.
When the perturbative approach is
carried out to the first order,
one finds
\be
\om^2\simeq
\frac{k}{m}\left(1-\frac{k}{K}\right).
\ee
This is a better approximation than (\ref{badone})
to the real natural frequency.
In this case we can even ask what will we
get by carrying out the perturbative approach
to all orders.
It is straightforward to see that the answer is
\be
\om^2=\frac{1}{2a}(b-\sqrt{b^2-4ac}).
\ee
This is precisely one of the two normal mode frequencies
of the exact equation of motion.
Although it does not tell us anything about
the other natrual frequency at higher energy
$\om^2\simeq\frac{K}{M}$,
it describes the exact effect of the stiff spring on
the low energy mode.

In this example, we see that the physical meaning
of reducing the phase space to the space of $q,\dot{q}$
in the perturbative formulation is to suppress
the excitation of high energy modes
in a way consistent with the exact evolution,
and only to allow low energy modes to be excited.

\subsection{SpaceTime NonCommutative Field Theory} \label{STNC}

Spacetime noncommutative field theories are
interesting and important for many reasons \cite{SST2}.
Although the naive quantization is
non-unitary and non-causal \cite{SST2},
perhaps a proper treatment will lead to
the noncommutative open string (NCOS) theory
\cite{SST2,GMMS}.

The exact treatment of this system is outside
the scope of this paper.
We will only study this theory perturbatively.
The zeroth order results are in fact already
studied in \cite{AB},
where it was shown that the kinetic term
with or without the $\ast$-product results
in the same Hamiltonian formulation.

Consider the Lagrangian
\be
L=\frac{1}{2}(\del_{\mu}\phi\del_{\mu}\phi-m^2\phi^2)
+\lam \phi\ast\phi\ast\phi.
\ee
The $\ast$-product is defined by
\be \label{star}
f\ast g(x)=e^{\frac{1}{2}\th^{\mu\nu}\del_{\mu}\del'_{\nu}}
f(x)g(x')|_{x'=x}.
\ee
It makes no difference whether we put $\ast$-products
in the quadratic terms.

For simplicity, consider the 1+1 dimensional case
\be
\label{tx}
[t,x]_{\ast}=\th.
\ee
As we saw in Sec. \ref{FirstOrder},
by a field redefinition,
the effective Lagrangian is given by (\ref{LL})
at the lowest order.
For the case at hand it is
\be \label{LLL}
\tilde{L}\simeq\frac{1}{2}
(\del_{\mu}\Phi\del_{\mu}\Phi-m^2\Phi^2)
+\lam \Phi\bast\Phi\bast\Phi,
\ee
where the $\bast$-product is defined as $\ast$
in (\ref{star}) with the replacement
\be
\del_t^n\rightarrow \left\{\begin{array}{ll}
         (-\del_x^2+m^2)^{n/2} & (n=\mbox{even}), \\
         (-\del_x^2+m^2)^{(n-1)/2}\del_t & (n=\mbox{odd}).
       \end{array}\right. \label{delt}
\ee
In the action defined by (\ref{LLL}),
one only needs $\del_t^n$-terms for even $n$.

%

This result clarifies a conceptual puzzle.
On one hand, the uncertainty relation
\be \label{uncer}
\Delta t\Delta x\geq |\th|,
\ee
which follows from (\ref{tx}),
implies that when we treat $t$ as a classical number,
$\Delta x\rightarrow \infty$.
Thus we should not associate
independent physical degrees of freedom
at different values of $x$.
It suggests a reduction of the number of degrees of freedom
\cite{HL,Li}.
On the other hand,
higher derivatives in the action
imply a larger phase space
in canonical formulation.
These two intuitions seem to contradict each other.

This puzzle may now be understood as follows.
The uncertainty relation (\ref{uncer})
does not really follow from (\ref{tx}),
because the time $t$, unlike $x$,
is not an operator in quantum mechanics.
However, according to (\ref{delt})
and the definition of $\bar{\ast}$-product,
it is an effective description
of the fact that the extent of spatial nonlocality
(the shift in $x$ for $\Phi$ in the interaction term)
is proportional to the energy scale,
which is approximated by the free energy
$(-\del_x^2+m^2)^{1/2}$ for small $\lam$.
Incidentally, this relation is reminicent of
the characteristic property of strings \cite{Yoneya}.
This is also one of the reasons why
spacetime noncommutative field theories
are of interest to us.
On the other hand,
in an exact treatment,
the consistent high energy completion
of spacetime noncommutative theory
is presumably the NCOS theory,
which contains a lot more degrees of freedom.

\section*{Acknowledgment}

The authors thank Marc Henneaux, Miao Li,
Feng-Li Lin, John Wang
for helpful discussions.
This work is supported in part by
the National Science Council,
the CosPA project of the Ministry of Education,
the National Center for Theoretical Sciences,
Taiwan, R.O.C.
and the Center for Theoretical Physics
at National Taiwan University.

\appendix

\section{One-Loop Unitarity for Derivative Interactions}

Consider a scalar field for example.
In $D$ dimensional spacetime,
its propagator is
\be
\Delta(x)=\th(x_0)\Delta^+(x)+\th(-x_0)\Delta^-(x),
\ee
where
\be
\Delta^{\pm}(x)=\frac{1}{(2\pi)^{D-1}}
\int d^D k \d(k^2+m^2)\th(\pm k_0)e^{ik\cdot x}.
\ee

{\center
\begin{picture}(300,140)(-150,-90) \label{Fig2}
\put(-65,-3){\hbox{\large Im}}
\put(0,0){\circle{40}}
\put(-20,0){\line(-1,0){20}}
\put(20,0){\line(1,0){20}}
\put(70,0){\makebox(0,0){=}}
\put(100,-30){
\begin{picture}(70,60)(-35,-30)
\put(-35,-25){\line(0,1){50}}
\put(0,0){\line(-1,0){30}}
\put(0,0){\line(1,1){20}}
\put(0,0){\line(1,-1){20}}
\put(30,-25){\line(0,1){50}}
\put(33,25){2}
\end{picture}
}
\put(40,-70){Figure 2}
\end{picture}
}

The left hand side of the diagram in Fig. 2 is
\be
\tilde{\Sigma}(x_f-x_i)=
ig^2 F(\del^{(1)}_f,\del^{(2)}_f)F(\del^{(1)}_i,\del^{(2)}_i)
\Delta_1(x^{(1)}_f-x^{(1)}_i)
\Delta_2(x^{(2)}_f-x^{(2)}_i) |_{x^{(1)}=x^{(2)}=x},
\ee
where the interaction is modified by
a (Hermitian) 
function $F$ of derivatives of two particles.
For the special case of interactions
in noncommutative field theories,
\be F(\del^{(1)},\del^{(2)})=
e^{\frac{i}{2}\th^{\mu\nu}\del^{(1)}_{\mu}\del^{(2)}_{\nu}}.
\ee
By Fourier transform,
\be
\Sigma(p)=i g^2 \int d^D x e^{-ip\cdot x}
F(\del)F(-\del)\Delta_1(x^{(1)})
\Delta_2(x^{(2)}) |_{x^{(1)}=x^{(2)}=x}.
\ee

If there is no time derivative in $F$,
we find
\bea
\mbox{Im}(\Sigma(p))&=&
\frac{g^2}{2}\int \frac{d^{D-1}k_1 d^{D-1}k_2}{(2\pi)^{D-2}}
\frac{F(ik_1,ik_2)}{2\om(k_1)}
\frac{F(-ik_1,-ik_2)}{2\om(k_2)}
\d^{D-1}(p-k_1-k_2)\cdot \nn\\
&& \cdot (\d(p_0-\om_1(k_1)-\om_2(k_2))+
\d(p_0+\om_1(k_1)+\om_2(k_2))),
\eea
which is exactly the statement of unitarity.
The same derivation can be extended to
other one-loop diagrams,
the point being that spatial coordinates
do not have to be explicitly integrated over
in the calculation.
Thus we see that interaction terms
with spatial derivatives preserve unitarity.
The derivation breaks down when
$F$ contains time derivatives.
The unitarity was found to be preserved for
like-like noncommutativity in \cite{light}.



\vskip .8cm
\baselineskip 22pt

\begin{thebibliography}{99}

\itemsep 0pt

\bibitem{BVDM}
A. O. Barvinsky, C. A. Vilkovisky,
Nucl. Phys. {\bf B282} (1987) 163;
{\bf B333} (1990) 471.

\bibitem{nonlocal}
A. Pais, G. E. Uhlenbeck, Phys. Rev. {\bf 79}, 145 (1950).

\bibitem{EW}
D.~A.~Eliezer and R.~P.~Woodard,
Nucl.\ Phys.\ B {\bf 325}, 389 (1989).

\bibitem{HH}
H. Hata,
Phys. Lett. {\bf B217} (1989) 438,
Nucl. Phys. {\bf B329} (1990) 698.

\bibitem{higherderiv}
K.~S.~Stelle,
Phys.\ Rev.\ D {\bf 16}, 953 (1977);
J.~Julve and M.~Tonin,
Nuovo Cim.\ B {\bf 46}, 137 (1978).

\bibitem{meson}
P. Kristensen, C. Moller, K. Dan. Vidensk.
Selsk. Mat-Fys. Medd. {\bf 27}, 7 (1952).


\bibitem{SW}
A.~Connes, M.~R.~Douglas and A.~Schwarz,
JHEP{\bf 9802}, 003 (1998)
[hep-th/9711162];
N.~Seiberg and E.~Witten,
JHEP{\bf 9909}, 032 (1999)
[hep-th/9908142].

\bibitem{SST1}
N.~Seiberg, L.~Susskind and N.~Toumbas,
JHEP {\bf 0006}, 044 (2000)
[hep-th/0005015].

\bibitem{Ostro}
M. Ostrogradski,
Mem. Ac. St. Petersbourg, {\bf VI 4} (1850) 385.

\bibitem{Weinb}
S. Weinberg, ``The Quantum Theory of Fields $I$'',
Sec. 3.1, 3.2 (1995), Cambridge.


\bibitem{BG1}
D. Barua, S. N. Gupta,
Phys. Rev. {\bf D16} (1977) 413.

\bibitem{JLM}
X. Ja\'{e}n, J. Llosa, A. Molina,
Phys. Rev. {\bf D34} (1986) 2302.

\bibitem{Simon}
J. Z. Simon,
Phys. Rev. {\bf D41} (1990) 3720;
J. Z. Simon,
Phys. Rev. {\bf D43} (1991) 3308.

\bibitem{FJ}
L.~D.~Faddeev and R.~Jackiw,
Phys.\ Rev.\ Lett.\  {\bf 60}, 1692 (1988).

\bibitem{HT}
M.~Henneaux and C.~Teitelboim,
``Quantization Of Gauge Systems,''
{\it  Princeton, USA: Univ. Pr. (1992)}.

\bibitem{LV}
J. Llosa, J. Vives,
J. Math. Phys. {\bf 35} (1994) 2856.

\bibitem{Woodard}
R.~P.~Woodard,
Phys.\ Rev.\ A {\bf 62}, 052105 (2000)
[hep-th/0006207].

\bibitem{GKL}
J.~Gomis, K.~Kamimura and J.~Llosa,
Phys.\ Rev.\ D {\bf 63}, 045003 (2001)
[hep-th/0006235].

\bibitem{Bering}
K.~Bering,
``A note on non-locality and Ostrogradski's construction,''
[hep-th/0007192].

\bibitem{BNW}
D.~L.~Bennett, H.~B.~Nielsen and R.~P.~Woodard,
Phys.\ Rev.\ D {\bf 57}, 1167 (1998)
[hep-th/9707088].

\bibitem{BR}
J.~L.~Barbon and E.~Rabinovici,
Phys.\ Lett.\ B {\bf 486}, 202 (2000)
[hep-th/0005073];
J.~Gomis and T.~Mehen,
Nucl.\ Phys.\ B {\bf 591}, 265 (2000)
[hep-th/0005129];
M.~Chaichian, A.~Demichev, P.~Presnajder and A.~Tureanu,
Eur.\ Phys.\ J.\ C {\bf 20}, 767 (2001)
[hep-th/0007156];
L.~Alvarez-Gaume, J.~L.~Barbon and R.~Zwicky,
JHEP {\bf 0105}, 057 (2001)
[hep-th/0103069];
T.~Mateos and A.~Moreno,
Phys.\ Rev.\ D {\bf 64}, 047703 (2001)
[hep-th/0104167];
A.~Bassetto, L.~Griguolo, G.~Nardelli and F.~Vian,
JHEP {\bf 0107}, 008 (2001)
[hep-th/0105257].

\bibitem{GW}
J.~Gomis and S.~Weinberg,
Nucl.\ Phys.\ B {\bf 469}, 473 (1996)
[hep-th/9510087].

\bibitem{SST2}
N.~Seiberg, L.~Susskind and N.~Toumbas,
JHEP {\bf 0006}, 021 (2000)
[hep-th/0005040].

\bibitem{GMMS}
R.~Gopakumar, J.~Maldacena, S.~Minwalla and A.~Strominger,
JHEP {\bf 0006}, 036 (2000)
[hep-th/0005048];
R.~Gopakumar, S.~Minwalla, N.~Seiberg and A.~Strominger,
JHEP {\bf 0008}, 008 (2000)
[hep-th/0006062].


\bibitem{AB}
R.~Amorim and J.~Barcelos-Neto,
``Remarks on the canonical quantization
of noncommutative theories,''
[hep-th/0108186].

\bibitem{HL}
P.~M.~Ho and M.~Li,
Nucl.\ Phys.\ B {\bf 596}, 259 (2001)
[hep-th/0004072];
P.~M.~Ho and M.~Li,
Nucl.\ Phys.\ B {\bf 590}, 198 (2000)
[hep-th/0005268].

\bibitem{Li}
M.~Li,
``Dimensional reduction via noncommutative spacetime:
Bootstrap and  holography,''
[hep-th/0103107].

\bibitem{Yoneya}
T. Yoneya, in ``Wandering in the Fields'',
eds. K. Kawarabayashi, A. Ukawa
(World Scientific, 1987) p. 419;
M. Li, T. Yoneya,
Phys. Rev. Lett. {\bf 78} (1997) 1219
[hep-th/9611072].

\bibitem{light}
O.~Aharony, J.~Gomis and T.~Mehen,
JHEP {\bf 0009}, 023 (2000)
[hep-th/0006236].










\end{thebibliography}
\end{document}